\documentclass[aps,pra,preprint,amssymb, amsfonts,amsmath,showpacs, nofootinbib,eqsecnum,superscriptaddress]{revtex4-1}
 \usepackage{graphicx}
  \usepackage{amssymb}
  \usepackage{color}

\begin{document}
\title{On the CSL Scalar Field Relativistic Collapse Model}
\author{Daniel Bedingham}
\email{daniel.bedingham@rhul.ac.uk}
\affiliation{Department of Physics, Royal Holloway, University of London, Egham, YW20 0EX.}
\author{Philip Pearle}
\email{ppearle@hamilton.edu}
\affiliation{Emeritus, Department of Physics, Hamilton College, Clinton, NY 13323}
\begin{abstract}
{The CSL dynamical collapse structure, adapted to the relativistically invariant model where the collapse-generating operator is a one-dimensional scalar field $\hat\phi(x,t)$ (mass $m$) is discussed.
A complete solution for the density matrix is given, for an initial state 
$|\psi,0\rangle=\frac{1}{\sqrt{2}}[|L\rangle+|R\rangle]$ when the Hamiltonian $\hat H$ is set equal to 0, and when $\hat H$  is the free field Hamiltonian. Here  $|L\rangle, |R\rangle$  are coherent states which represent clumps of particles, with mean particle number density $N\chi_{i}^{2}(x)$, 
where $\chi_{1}(x),\chi_{1}(x) $    
are gaussians of width $\sigma>>m^{-1}$ with mean positions separated by distance $>>\sigma$. It is shown that, with high probability, the solution for $\hat H=0$ (identical to 
the short time solution for $\hat H\neq 0$) favors collapse toward  eigenstates  of the scalar field 
whose eigenvalues are close to $\sim\chi_{i}(x)$.   Thus, this collapse dynamics results in essentially one clump of particles. 
However, eventually particle production dominates the density matrix since,
as is well known, the collapse generates energy/sec-volume of every particle momentum in equal amounts.  
Because of the particle production, this is not an experimentally viable physical theory but, as is emphasized by the discussion,  it is a sound relativistic collapse model, with sensible collapse behavior.}
 \end{abstract}

\maketitle

\section{Introduction}\label{I}

The non-relativistic Continuous  Spontaneous Localization (CSL) model\cite{P},\cite{GPR} is a response to the quantum measurement problem. It is a modification of standard quantum dynamics with nonlinear and stochastic features, designed to exhibit spontaneous collapse of the wavefunction. This happens in such a way that small systems are effectively unaffected, whilst large-scale superpositions of differing distributions of matter are rapidly suppressed. The dynamical structure of CSL can then describe both the unitary development of small quantum systems and the collapse which occurs during a quantum measurement. 

This is a specialization of a general CSL structure that describes dynamical collapse of the state vector toward an eigenstate of any set of commuting operators, 
termed `collapse-generating' operators. In the case of non-relativistic CSL, these are mass-density operators at each point of space (smeared over a characteristic distance). In this paper, we take the collapse-generating operators to be the relativistic scalar quantum field operator $\hat\phi(x,0)$ associated to particles of mass $m$ at each point of space.

The state vector in the Schr\"odinger picture and in the interaction picture are respectively
\begin{eqnarray}\label{1.1}
|\psi ,T\rangle_{S}&=&{\cal T} e^{-i\int_{0}^{T}dt\hat H   -\frac{1}{4\lambda}\int_{0}^{T}dxdt[w(x,t)-2\lambda \hat\phi(x,0)]^{2}}|\psi ,0\rangle,\nonumber\\
|\psi ,T\rangle_{I}&=&{\cal T} e^{-\frac{1}{4\lambda}\int_{0}^{T}dxdt[w(x,t)-2\lambda\hat \phi(x,t)]^{2}}|\psi ,0\rangle, 
\end{eqnarray} 
\noindent where ${\cal T}$ is the time-ordering operator,  $\hat H$ is the  Hamiltonian for the freely 
evolving scalar field,  $w(x,t)$ is a random real-valued scalar field of 
white noise type, and these state vectors (whose norms are not 1) are to occur in nature with probability proportional to the state vector squared norm.

It is known\cite{Pearle2}\cite{GGP} that this model generates problematic divergent energy increase and this will be demonstrated in Section \ref{S:3}. Realistic collapse models generally display some energy increase as a result of the fact that localisation is necessarily accompanied by some spreading in momentum. This offers the possibility of experimentally testing CSL via heating effects. Here we leave aside the issue of infinite energy increase, assuming that it can eventually be regulated in some way. Instead we focus on the collapse behaviour. For an initial state representing a superposition of two differently located clumps of particles we will find that the state collapses toward an eigenstate of the scalar field representing either one of the two clumps of particles.

The density matrix in the Schr\"odinger picture satisfies the Lindblad evolution  equation  
\begin{eqnarray}\label{1.2}
\frac{d}{dt}\hat\rho(t)
&=&-i[\hat H,\hat\rho(t)]-\frac{\lambda}{2}\int dx[\hat\phi(x,0)[\hat\phi(x,0),\hat\rho(t)]].                                    
\end{eqnarray}
We will now show how the density matrix can be constructed as the direct product of harmonic oscillator density matrices associated to each momentum mode (this is similar, but not identical to, the 
construction in \cite{Pearle2}). 

Write the particle annihilation operator as $\hat a(k)\equiv\hat a_{k}/\sqrt{dk}$, so $[\hat a_{k},\hat a^{\dagger}_{k'}]=\delta_{k,k'}$. Next, define position and momentum operators 
associated to each momentum mode (of course, these have nothing to do with position and momentum for the actual particles), for $k>0$, 
$\hat a_{k}\equiv\frac{1}{\sqrt{2}}[\hat x_{k1}+i\hat p_{k1}]$ and, 
for $k<0$, $\hat a_{-|k|}\equiv \hat b_{|k|}=\frac{1}{\sqrt{2}}[\hat x_{|k|2}+i \hat p_{|k|2}]$. Finally define center of mass and relative position and momentum operators for only $k>0$, 
$\hat X_{k}\equiv\frac{1}{2}[\hat x_{k1}+\hat x_{k2}],\hat P_{k}\equiv[\hat p_{k1}+\hat p_{k2}],  
\hat x_{k}\equiv[\hat x_{k1}-\hat x_{k2}], \hat p_{k}\equiv\frac{1}{2}[\hat p_{k1}-\hat p_{k2}]$. 

The scalar field $\hat\phi(x,0)=\int_{-\infty}^{\infty}dk\sqrt{\frac{1}{4\pi\omega(k)}}[\hat a(k)e^{ikx} + \hat a^{\dagger}(k)e^{-ikx}]$ may then be written as
\begin{eqnarray}\label{1.3}
\hat\phi(x,0)&=&\int_{0}^{\infty}dk\sqrt{\frac{1}{4\pi\omega(k)}}[(\hat a(k)+\hat b^{\dagger}(k))e^{ikx} +( \hat a^{\dagger}(k)+\hat b(k))e^{-ikx}]\nonumber\\
  &=&\sum_{k>0}\sqrt{\frac{dk}{2\pi\omega(k)}}[ (\hat X_{k}+i\hat p_{k})e^{ikx} + (\hat X_{k}-i\hat p_{k})e^{-ikx}].                         
\end{eqnarray}

Define eigenvectors of the center of mass position and relative momentum operators, $|X\rangle|p\rangle$, where $\hat X_{k}|X\rangle|p\rangle=X_{k}|X\rangle|p\rangle,\; \hat p_{k}|X\rangle|p\rangle=p_{k}|X\rangle|p\rangle$, so   
$|X\rangle|p\rangle=\prod_{k>0}|X_{k}\rangle|p_{k}\rangle$, and $\langle X'_{k'}|X_{k}\rangle=\delta_{kk'}\delta (X'_{k}-X_{k}), \;\langle p'_{k'}|p_{k}\rangle=\delta_{kk'}\delta (p'_{k}-p_{k})$. 

Then, since $[\hat\phi(x,0),\hat\phi(x',0)]=0$, there is a joint eigenvector (all $x$) satisfying $\hat\phi(x,0)|f\rangle\equiv f(x)|f\rangle$, with real arbitrary eigenvalue functions $f(x)=\frac{1}{\sqrt{2\pi}}\int_{-\infty}^{\infty}dk\tilde f(k) e^{ikx}$ (where $\tilde f^{*}(k)=\tilde f(-k)$).  It follows from (\ref{1.3}) that $|X\rangle|p\rangle$ is an eigenstate of  $\hat\phi(x,0)$, 
\begin{eqnarray}\label{1.4}
\hat\phi(x,0)|X\rangle|p\rangle 
  &=&\sum_{k>0}\sqrt{\frac{dk}{2\pi\omega(k)}}[ (X_{k}+ip_{k})e^{ikx} + ( X_{k}-ip_{k})e^{-ikx})]|X\rangle|p\rangle,                         
\end{eqnarray}
\noindent so, with $\tilde f(k)=\tilde f_{R}(k)+i\tilde f_{I}(k)$, we identify $\frac{1}{\sqrt{dk\omega(k)}}X_{k}=\tilde f_{R}(k),\frac{1}{\sqrt{dk\omega(k)}}p_{k}=\tilde f_{I}(k)$,
and then
\begin{eqnarray}\label{1.5}
|f\rangle &=&\prod_{k>0}|\sqrt{\omega(k) dk}\tilde f_{R}(k) \rangle|\sqrt{\omega dk} \tilde f_{I}(k) \rangle\equiv \prod_{k>0}|f\rangle_{k}.                     
\end{eqnarray}
\noindent According to  (\ref{1.5}),  $_{k}\langle f'|f\rangle_{k}=\frac{1}{\omega(k) dk}\delta(f'_{R}(k)-f_{R}(k))\delta(f'_{I}(k)-f_{I}(k))$.

 By inserting (\ref{1.3}) into (\ref{1.2}), the Lindblad equation becomes 
 \begin{eqnarray}\label{1.6}
\frac{d}{dt}\hat\rho(t)
&=&\sum_{k>0}\Big\{-i\omega(k)[\hat X^{2}_{k}+\frac{1}{4}\hat P^{2}_{k},\hat\rho(t)]-\frac{\lambda}{\omega(k)}[\hat X_{k}, [\hat X_{k}, \hat\rho(t)]]            \nonumber\\
&& -i\omega(k)[ \frac{1}{4}\hat x^{2}_{k}+\hat p^{2}_{k},\hat\rho(t)] -\frac{\lambda}{\omega(k)} [\hat p_{k}, [\hat p_{k}, \hat\rho(t)]]\Big\}.                                
\end{eqnarray}
The problem can be reduced to solving individual mode equations if the initial density matrix can be written as $\hat\rho(0)=\sum_{\nu}c_{\nu}\prod_{k}\hat\rho_{\nu,k}(0)$. This defines a state whose modes are separable. For initial states of this type 
the density matrix at later times may be written as $\hat\rho(t)=\sum_{\nu}c_{\nu}\prod_{k}\hat\rho_{\nu,k}(t)$ where 
 \begin{eqnarray}\label{1.7}
\frac{d}{dt}\hat\rho_{\nu,k}(t)
&=&-i\omega(k)[\hat X^{2}_{k}+\frac{1}{4}\hat P^{2}_{k},\hat\rho_{\nu,k}(t)]-\frac{\lambda}{\omega(k)}[\hat X_{k}, [\hat X_{k}, \hat\rho_{\nu,k}(t)]]            \nonumber\\
&& -i\omega(k)[ \frac{1}{4}\hat x^{2}_{k}+\hat p^{2}_{k},\hat\rho_{\nu,k}(t)] -\frac{\lambda}{\omega(k)} [\hat p_{k}, [\hat p_{k}, \hat\rho_{\nu,k}(t)]],                                
\end{eqnarray}
\noindent subject to the initial condition $\hat\rho_{\nu,k}(0)$.

\section{The Initial State}

We will be considering the initial state 
\begin{equation}\label{2.1}
|\psi,0\rangle=\frac{1}{\sqrt{2[1+\langle \ell_{1}|\ell_{2}\rangle] }} [|\ell_{1}\rangle+|\ell_{2}\rangle],\quad \hat\rho(0)=
\frac{1}{2[1+\langle \ell_{1}|\ell_{2}\rangle]}[|\ell_{1}\rangle+|\ell_{2}\rangle][\langle\ell_{1}|+\langle\ell_{2}|],
\end{equation}
\noindent where $|\ell_{1}\rangle, |\ell_{2}\rangle$ represent two clumps of particles at widely separated locations: 
\begin{eqnarray}\label{2.2}
|\ell\rangle&\equiv &e^{\int_{-\infty}^{\infty}dk\hat a^{\dagger}(k)\tilde\chi(k)}|0\rangle e^{-\frac{1}{2}\int_{-\infty}^{\infty}dk |\tilde\chi(k)|^{2}    }\nonumber\\
&=&\prod_{k>0}e^{\sqrt{2dk}[(\hat X_{k}-i\frac{1}{2}\hat P_{k})\tilde\chi_{R}(k) + (\hat p_{k}+i\frac{1}{2}\hat x_{k})\tilde\chi_{I}(k)]}|0\rangle_{k} e^{-dk |\tilde\chi(k)|^{2}    }\equiv \prod_{k>0}|\ell\rangle_{k} ,
\end{eqnarray}
\noindent with
\begin{eqnarray}\label{2.3}
\chi(x)&\equiv &N^{1/2}\frac{1}{(2\pi\sigma^{2})^{1/4}} e^{-\frac{1}{4\sigma^{2}}(x-\ell)^{2}} =\frac{1}{\sqrt{2\pi}}\int_{-\infty}^{\infty}dk\tilde\chi(k)e^{ikx}, \nonumber\\
  \tilde\chi(k)&=&N^{1/2}\Big(\frac{2\sigma^{2}}{\pi} \Big)^{1/4} e^{-k^{2}\sigma^{2}}e^{-ik\ell}.                         
\end{eqnarray}
\noindent and $|0\rangle=\prod_{k>0}|0\rangle_{k}$ is the vacuum (no particle) state,  i.e.~$|l\rangle$ is a form of coherent state.  

The particle number density operator is $\hat\xi^{\dagger}(x)\hat\xi(x)$ (where $\hat\xi(x)=\frac{1}{\sqrt{2\pi}}\int_{-\infty}^{\infty}dk \hat a(k)e^{ikx})$.  Since  $\langle \ell| \hat\xi^{\dagger}(x)\hat\xi(x)|\ell\rangle=\chi^{2}(x)$, this state may be thought of as representing 
$\approx N$ particles (the particle number operator $\hat N= \int_{-\infty}^{\infty}dx\hat\xi^{\dagger}(x)\hat\xi(x)$ has mean value $N$ and mean-squared value $N^{2}+N$, so 
the standard deviation of the number of particles divided by $N$ is $1/\sqrt{N}$) centered  at $\ell$ and spread over width $\approx \sigma<<|\ell_{1}-\ell_{2}|$.

The states $|\ell_{1}\rangle, |\ell_{2}\rangle$ aren't quite orthogonal:
\begin{eqnarray}\label{2.4}
\langle\ell_{1}|\ell_{2}\rangle&=&e^{\int_{-\infty}^{\infty} dk\tilde\chi_{1}^{*}(k)\tilde\chi_{2}(k)}e^{-\frac{1}{2}\int_{-\infty}^{\infty}dk |\tilde\chi_{1}(k)|^{2}}
e^{-\frac{1}{2}\int_{-\infty}^{\infty}dk |\tilde\chi_{2}(k)|^{2}}\nonumber\\
&=&
e^{-N\big[1-e^{-\frac{(\ell_{1}-\ell_{2})^{2}}{8\sigma^{2}}}\big]}.                    
\end{eqnarray}
However, we will assume the number of particles $N$ is so large that we may neglect $e^{-N}$  compared to 1, in sections IV, V.   Thus, we will take the state vector normalization factor in 
(\ref{2.1}) to be simply $\frac{1}{\sqrt{2}}$. Therefore, we have to solve Eq.(\ref{1.7}) for 
\begin{equation}\label{2.5}
\hat\rho_{11k}(t), \hat\rho_{12k}(t), \hat\rho_{21k}(t), \hat\rho_{22k}(t), \hbox{ with corresponding initial conditions } \hat\rho_{ss'k}(0)=|\ell_{s}\rangle_{k}\negthinspace_{k}\langle \ell_{s'}|.
\end{equation}

We will need to know $|0\rangle_{k}$ and  $|\ell\rangle_{k}$ in the $|X_{k}\rangle |p_{k}\rangle$ basis. 

For $|0\rangle_{k}$, since 
$(\hat X_{k}+i\frac{1}{2}\hat P_{k})|0\rangle_{k}=0,\; (\hat p_{k}-i\frac{1}{2}\hat x_{k})|0\rangle_{k}=0$: 
\begin{eqnarray}\label{2.6}
\langle X_{k}|\langle p_{k}|0\rangle_{k}=\sqrt{\frac{2}{\pi}}e^{-X_{k}^{2}}  e^{-p_{k}^{2}}.                     
\end{eqnarray}

For $|\ell\rangle_{k}$, we apply the Campbell-Baker-Hausdorff theorem to see that 
\begin{eqnarray}\label{2.7}
e^{\alpha(X-\frac{1}{2}\frac{d}{dX})} e^{-X^{2}} e^{-\frac{1}{2}\alpha^{2}}&=& e^{\alpha X} e^{-\alpha\frac{1}{2}\frac{d}{dX}}e^{-X^{2}} e^{-\frac{3}{4}\alpha^{2}} \nonumber\\
 &=& e^{\alpha X}e^{-(X-\frac{1}{2}\alpha)^{2}} e^{-\frac{3}{4}\alpha^{2}}=e^{-(X-\alpha)^{2}}.        
\end{eqnarray}
\noindent Therefore, it follows from (\ref{2.2}),  (\ref{2.7}) that
\begin{eqnarray}\label{2.8}
\langle X_{k}|\langle p_{k}|\ell\rangle_{k}=\sqrt{\frac{2}{\pi}}e^{-[X_{k}-\sqrt{2dk}\tilde\chi_{R}(k)]^{2}}e^{-[p_{k}-\sqrt{2dk}\tilde\chi_{I}(k)]^{2}},       
\end{eqnarray}
or, using (\ref{1.5}),
\begin{eqnarray}\label{2.9}
\negthinspace_{k}\langle f|\ell\rangle_{k}&=&\sqrt{\frac{2}{\pi}}e^{-dk[\sqrt{\omega(k)}\tilde f_{R}(k)-\sqrt{2}\tilde\chi_{R}(k)]^{2}}e^{-dk[\sqrt{\omega(k)}\tilde f_{I}(k)-\sqrt{2}\tilde\chi_{I}(k)]^{2}} \nonumber\\
&\approx&\sqrt{\frac{2}{\pi}}e^{-dk\omega(k)[\tilde f_{R}(k)-\sqrt{\frac{2}{m}}\tilde\chi_{R}(k)]^{2}}e^{-dk\omega(k)[\tilde f_{I}(k)-\sqrt{\frac{2}{m}}\tilde\chi_{I}(k)]^{2}},
\end{eqnarray}
\noindent where we have made the approximation $\frac{1}{\sqrt{\omega(k)}}\approx\frac{1}{\sqrt{m}}$ since $\tilde\chi(k)$ is non-negligible only for $k<<m$.

\section{Particle Production}
\label{S:3}

The eigenstates of the scalar field $|f\rangle$ are superpositions of states with all numbers of particles.  To see this, we note that  an alternative form for $|f\rangle$ is
\begin{eqnarray}\label{3.1}
|f\rangle&\sim&e^{-\int_{0}^{\infty}dka^{\dagger}(k)b^{\dagger}(k)} e^{\int_{0}^{\infty}dk\sqrt{2\omega}[a^{\dagger}(k)\tilde f(k) +b^{\dagger}(k)\tilde f^{*}(k)]}|0\rangle\nonumber\\
&\sim&e^{-\sum_{k>0}a^{\dagger}_{k}b^{\dagger}_{k}} e^{  \sum_{k>0}\sqrt{2\omega dk}[a^{\dagger}_{k}\tilde f(k)+b^{\dagger}_{k}\tilde f^{*}(k)]}|0\rangle. 
\end{eqnarray}
\noindent This is because the right side of (\ref{3.1}) is an eigenstate of $\hat\phi(x,0)$ with eigenvalue $f(x)$:
\begin{eqnarray}\label{3.2}
&&\hat\phi(x,0)e^{-\sum_{k>0}a^{\dagger}_{k}b^{\dagger}_{k}} e^{  \sum_{k>0}\sqrt{2\omega dk}[a^{\dagger}_{k}\tilde f(k)+b^{\dagger}_{k}\tilde f^{*}(k)]}|0\rangle\nonumber\\
&=&\sum_{k>0}\sqrt{\frac{dk}{4\pi\omega(k)}}[(\hat a_{k}+\hat b^{\dagger}_{k})e^{ikx} +( \hat a^{\dagger}_{k}+\hat b_{k})e^{-ikx}]
e^{-\sum_{k>0}a^{\dagger}_{k}b^{\dagger}_{k}} e^{  \sum_{k>0}\sqrt{2\omega dk}[a^{\dagger}_{k}\tilde f(k)+b^{\dagger}_{k}\tilde f^{*}(k)]}|0\rangle\nonumber\\
&=&e^{-\sum_{k>0}a^{\dagger}_{k}b^{\dagger}_{k}}\sum_{k>0}\sqrt{\frac{dk}{4\pi\omega(k)}}[\hat a_{k}e^{ikx} + \hat b_{k}e^{-ikx}]
e^{  \sum_{k>0}\sqrt{2\omega dk}[\hat a^{\dagger}_{k}\tilde f(k)+ \hat b^{\dagger}_{k}\tilde f^{*}(k)]}|0\rangle\nonumber\\
 &=&e^{-\sum_{k>0}\hat a^{\dagger}_{k} \hat b^{\dagger}_{k}}e^{  \sum_{k>0}\sqrt{2\omega dk}[\hat a^{\dagger}_{k}\tilde f(k)+
 \hat b^{\dagger}_{k}\tilde f^{*}(k)]}|0\rangle\sum_{k>0}dk\sqrt{\frac{1}{2\pi}}[\tilde f_{R}(k)e^{ikx} +\tilde f_{I}(k)e^{-ikx}]
 \nonumber\\
 &=&e^{-\sum_{k>0}\hat a^{\dagger}_{k} \hat b^{\dagger}_{k}}e^{  \sum_{k>0}\sqrt{2\omega dk}[a^{\dagger}_{k}\tilde f(k)+b^{\dagger}_{k}\tilde f^{*}(k)]}|0\rangle f(x).
 \end{eqnarray}

For example, the state $|f=0\rangle$ may be written as
\begin{eqnarray}\label{3.3}
|f=0\rangle&\sim&e^{-\sum_{k>0}\hat a^{\dagger}_{k}\hat b^{\dagger}_{k}}|0\rangle=\prod_{k} \sum_{n=0}^{\infty}(-1)^{n}\frac{1}{n!}(\hat a^{\dagger}_{k}\hat b^{\dagger}_{k})^{n}|0\rangle_{k}\nonumber\\
&=&\prod_{k}\sum_{n=0}^{\infty}(-1)^{n}|n\rangle_{k}|n\rangle_{-k}.
\end{eqnarray}
\noindent For $f(x)\neq 0$, this is multiplied by a factor which puts even more particles into each mode. 

Since any  eigenstate of the scalar field has infinite particles, and the collapse end product is one or another of these eigenstates,  the 
collapse process must generate these particles. We calculate the rate of particle production (this is well known\cite{P}, but repeated here for completeness, in the formalism of this paper).  It follows from (\ref{1.6}) that 
\begin{eqnarray}\label{3.4}
\frac{d}{dt}\overline {\hat O}\equiv \frac{d}{dt}\hbox{Tr}\hat O\hat\rho(t)
&=&\sum_{k>0}\Big\{i\omega(k)\hbox{Tr}\hat\rho(t)[\hat a^{\dagger}_{k}\hat a_{k}+\hat b^{\dagger}_{k}\hat b_{k},\hat O ] \nonumber\\
&&-\frac{\lambda}{\omega(k)}\hbox{Tr}\rho(t)[\hat X_{k}, [\hat X_{k}, \hat O] ]          
 -\frac{\lambda}{\omega(k)}\rho(t) [\hat p_{k}, [\hat p_{k}, \hat O]]\Big\}.                                
\end{eqnarray}
\noindent where $\hbox{Tr}$ is the trace operation. 
Setting $\hat O$ equal to
\begin{eqnarray}\label{3.5} 
\hat a^{\dagger}_{k}\hat a_{k}+\hat b^{\dagger}_{k}\hat b_{k}&=&\hat X_{k}^{2}+\frac{1}{4}\hat P_{k}^{2}+\hat p_{k}^{2}+\frac{1}{4}\hat x_{k}^{2}-1 \hbox{ and }
\hat a^{\dagger}_{k}\hat a_{k}-\hat b^{\dagger}_{k}\hat b_{k}=X_{k}x_{k}+P_{k}p_{k},
\end{eqnarray}
\noindent and inserting this into (\ref{3.4}), we find $\frac{d}{dt}\overline{\hat a^{\dagger}_{k}\hat a_{k}+\hat b^{\dagger}_{k}\hat b_{k}}= \frac{\lambda }{\omega(k)}$, 
$\frac{d}{dt}\overline{\hat a^{\dagger}_{k}\hat a_{k}-\hat b^{\dagger}_{k}\hat b_{k}}= 0$ and so
\begin{equation}\label{3.6} 
\overline{\hat a^{\dagger}_{k}\hat a_{k}(t)}= \frac{\lambda t}{2\omega(k)}+ \overline{\hat a^{\dagger}_{k}\hat a_{k}(0)}, \;\overline{\hat b^{\dagger}_{k}\hat b_{k}(t)}=\frac{\lambda t}{2\omega(k)}+ \overline{\hat b^{\dagger}_{k}\hat b_{k}(0)}. 
\end{equation}

In particular, each momentum mode contributes the same rate of energy increase,
\begin{equation}\label{3.7} 
\overline{\hat H_{k}(t)}=\omega(k)[\overline{\hat a^{\dagger}_{k}\hat a_{k}+\hat b^{\dagger}_{k}\hat b_{k}}]=\lambda t+ \overline{\hat H_{k}(0)}.
\end{equation} 
To clarify what that means, we use $dk=2\pi/\int dx$, which follows from
\begin{equation}\label{3.8} 
[\hat a_{k},\hat a_{k}^{\dagger}]=1=dk[\hat a(k), \hat a^{\dagger}(k)]=dk\delta(k-k)=dk\frac{1}{2\pi}\int dxe^{ix(k-k)}=dk\frac{1}{2\pi}\int dx.
\end{equation} 
\noindent Replacing $\int dx$ by the length $\equiv L$, and Inserting this into (\ref{3.7}):
 \begin{equation}\label{3.9} 
\frac{\overline{\hat H_{k}(t)}}{L}
=\frac{dk}{2\pi}\lambda t+ \frac{\overline{\hat H_{k}(0)}}{L}.
\end{equation}
That is, the energy per unit length contributed by each mode of $|$momentum$|=k>0$, of width $dk$, grows linearly with time. The net contribution of a finite range of modes is finite, but the net contribution from all modes is infinite at any finite time.  And, at infinite time, each mode has infinite energy/length, which conforms with 
the end collapse to field eigenstates which have an infinite number of particles in each mode.

\section{Collapse Of A Superposition When $H=0$}

The initial density matrix is to be $\hat \rho(0)=\frac{1}{2}[|\ell_{1}\rangle+\ell_{2}\rangle][\langle \ell_{1}|+\langle \ell_{2}|]$, corresponding to a superposition of $\approx N$ particles in two 
widely separated clumps,

When $H=0$, the solution of the Lindblad Eq.(\ref{1.7}), in the $|X_{k}\rangle|p_{k}\rangle$ basis and in the $|f\rangle_{k}$ basis is, using (\ref{2.8}), (\ref{2.9}),  
\begin{eqnarray}\label{4.1}
\langle X_{k}|\langle p_{k}|\hat\rho_{ss' k} (t)|X'_{k}\rangle|p'_{k}\rangle&=&e^{-\frac{\lambda t}{\omega(k)}\{[X_{k}-X'_{k}]^{2}+[p_{k}-p'_{k}]^{2}\}  }\nonumber\\
&&\frac{2}{\pi}e^{-[X_{k}-\sqrt{2dk}\tilde\chi_{Rs}(k)]^{2}-[p_{k}-\sqrt{2dk}\tilde\chi_{Is}(k)]^{2} }
e^{-[X'_{k}-\sqrt{2dk}\tilde\chi_{Rs'}(k)]^{2}-[p'_{k}-\sqrt{2dk}\tilde\chi_{Is'}(k)]^{2} },\nonumber\\
_{k}\langle f|\hat\rho_{ss' k} (t)|f'\rangle_{k}&=&e^{-\lambda tdk|\tilde f(k)-\tilde f'(k)|^{2}}
\frac{2}{\pi}e^{-dk\omega(k)|\tilde f(k)-\sqrt{\frac{2}{m}}\tilde\chi_{s}(k)|^{2}}e^{-dk\omega(k)|\tilde f'(k)-\sqrt{\frac{2}{m}}\tilde\chi_{s'}(k)|^{2}}
\end{eqnarray}
\noindent Using (\ref{4.1}), we then can construct the complete density matrix in the $|f\rangle$ basis:
\begin{eqnarray}\label{4.2}
\langle f|\hat\rho(t)|f'\rangle&=&\frac{1}{2}e^{-\lambda t\int_{0}^{\infty }dk|\tilde f(k)-\tilde f'(k)|^{2}}\sum_{ss'=1}^{2}
e^{-\int_{0}^{\infty }dk\omega(k)|\tilde f(k)-\sqrt{\frac{2}{m}}\tilde\chi_{s}(k)|^{2}}e^{-\int_{0}^{\infty }dk\omega(k)|\tilde f'(k)-\sqrt{\frac{2}{m}}\tilde\chi_{s'}(k)|^{2}}\nonumber\\
&=&\frac{1}{2}e^{-\lambda t\frac{1}{2}\int_{-\infty}^{\infty }dx\big[ f(x)- f'(x)\big]^{2}}\nonumber\\
&&
\sum_{ss'=1}^{2}e^{-\frac{1}{2}\int_{-\infty}^{\infty }dx\big\{\big[m^{2}-d^{2}/dx^{2}\big]^{1/4}\big[f(x)-\sqrt{\frac{2}{m}}\chi_{s}(x)\big]\big\}^{2}}e^{-\frac{1}{2}\int_{-\infty}^{\infty }dx\big\{\big[m^{2}-d^{2}/dx^{2}\big]^{1/4}\big[f'(x)-\sqrt{\frac{2}{m}}\chi_{s'}(x)\big]\big\}^{2}}
\nonumber\\
\end{eqnarray}
\noindent using Parseval's theorem, and absorbing the factors $\frac{2}{\pi}$ in the functional integral element $Df\equiv\prod_{k>0}d\tilde f_{R}(k)d\tilde f_{I}(k)\frac{2dk\omega(k)}{\pi}$.

The first exponential in Eqs.(\ref{4.2})  is largest when $f(x)\approx f'(x)$: this term is responsible for the ultimate ($t\rightarrow\infty$) collapse to eigenstates of the scalar field.  

The exponential terms in the sum are largest when $f(x)\approx \sqrt{2/m}\chi_{s}(x), f'(x)\approx \sqrt{2/m}\chi_{s'}(x)$. For the  contribution to the density matrix of 
the diagonal elements of the initial density matrix,  $s=s'$, this is compatible with $ f(x)-f'(x)\approx 0$, and so 
 there can be relatively little decay. 
 
 However, for the terms associated with the off-diagonal elements of the initial density matrix, $s\neq s'$, the conditions  $f(x)\approx f'(x)$,   
 $f(x)\approx \sqrt{2/m}\chi_{s}(x)$,   $f'(x)\approx \sqrt{2/m}\chi_{s'}(x)$ cannot all be satisfied, and so these terms exponentially decay as time progresses. 

To see this is more detail we can find the values of $\tilde{f}(k)$ and $\tilde{f}'(k)$ which maximise the exponent in (\ref{4.1}). These are 
\begin{eqnarray}
 \tilde{f}_0(k) &=& \frac{1}{2\lambda t  + \omega}\sqrt{\frac{2}{m}}\left[\lambda t (\tilde{\chi}_s(k) +\tilde{\chi}_{s'}(k)) + \omega\tilde{\chi}_{s}(k)\right], \nonumber\\
\tilde{f}_0'(k) &=& \frac{1}{2\lambda t  + \omega}\sqrt{\frac{2}{m}}\left[\lambda t (\tilde{\chi}_s(k) +\tilde{\chi}_{s'}(k)) + \omega\tilde{\chi}_{s'}(k)\right].
\end{eqnarray}
When $\tilde{f}$ and $\tilde{f}'$ take these forms, the exponent in (\ref{4.1}) takes the value
\begin{equation}
-\frac{2\lambda t\omega}{m(2\lambda t + \omega)}|\tilde{\chi}_s(k) - \tilde{\chi}_{s'}(k) |^2.
\end{equation}
Therefore, in the limit that $t\rightarrow \infty$, if $s = s'$, the exponent is a maximum with value 0 when $\tilde{f}(k)  = \tilde{f}'(k)= \sqrt{2/m}\tilde{\chi}_s(k)$; and, if $s\neq s'$, the exponent is a maximum with value $-(\omega/m)|\tilde{\chi}_s - \tilde{\chi}_{s'} |^2 \approx -|\tilde{\chi}_s - \tilde{\chi}_{s'} |^2 $ when $\tilde{f}(k) = \tilde{f}'(k) = \sqrt{1/2m}(\tilde{\chi}_s(k) +\tilde{\chi}_{s'}(k))$. In the latter case, taking into account all modes,
\begin{equation}
\lim_{t\rightarrow \infty}\langle f_0 |\rho_{ss'} | f_0'\rangle = e^{-\int_0^{\infty}dk |\tilde{\chi}_s - \tilde{\chi}_{s'} |^2} = e^{-N\big[1-e^{-\frac{(\ell_{1}-\ell_{2})^{2}}{8\sigma^{2}}}\big]},
\end{equation}
where we have used (\ref{2.4}). This represents the overlap between states $|l_1\rangle$ and $|l_2\rangle$ which for large $N$ is negligibly small. 
This reflects the small probability at large $t$ of the particles belonging to both clumps.

  In non-relativistic CSL the collapse, in a superposition  of two clumps of matter, toward one or another clump 
 occurs because the clumps represent two quite different mass-density distributions. 
Here, the collapse occurs because the two clumps correspond to two quite different scalar field distributions which are approximately proportional to 
the two separated clump wave functions representing  two quite different mass-density distributions.

Last, lets look at  the density matrix's components in the initial clump state basis. Using (\ref{2.8}) and (\ref{4.1}), 
 \begin{eqnarray}\label{4.3}
_{k}\langle \ell_{i}|\hat \rho_{ss' k}(t)|\ell_{j}\rangle_{k}&=&\int dX_{k}dp_{k}\int dX'_{k}dp'_{k}e^{-\frac{\lambda t}{\omega(k)}\{[X_{k}-X'_{k}]^{2}+[p_{k}-p'_{k}]^{2}\}  }\nonumber\\
&&_{k}\langle \ell_{i}|X_{k}\rangle|p_{k}\rangle \langle X_{k}|\langle p_{k}|\ell_{s}\rangle_{k k}\langle\ell_{s'}|X'_{k}\rangle|p'_{k}\rangle\
\langle X'_{k}|\langle p'_{k}|\ell_{j}\rangle_{k}\nonumber\\
&=&\int dX_{k}dp_{k}\int dX'_{k}dp'_{k}e^{-\frac{\lambda t}{\omega(k)}\{[X_{k}-X'_{k}]^{2}+[p_{k}-p'_{k}]^{2}\}  }\nonumber\\
&&\sqrt{\frac{2}{\pi}}e^{-[X_{k}-\sqrt{2dk}\tilde\chi_{iR}(k)]^{2}}e^{-[p_{k}-\sqrt{2dk}\tilde\chi_{iI}(k)]^{2}}\sqrt{\frac{2}{\pi}}
e^{-[X_{k}-\sqrt{2dk}\tilde\chi_{sR}(k)]^{2}}e^{-[p_{k}-\sqrt{2dk}\tilde\chi_{sI}(k)]^{2}}\nonumber\\
&&\sqrt{\frac{2}{\pi}}e^{-[X'_{k}-\sqrt{2dk}\tilde\chi_{jR}(k)]^{2}}e^{-[p'_{k}-\sqrt{2dk}\tilde\chi_{jI}(k)]^{2}}\sqrt{\frac{2}{\pi}}
e^{-[X'_{k}-\sqrt{2dk}\tilde\chi_{s'R}(k)]^{2}}e^{-[p'_{k}-\sqrt{2dk}\tilde\chi_{s'I}(k)]^{2}}.\nonumber\\
\end{eqnarray}
\noindent We evaluate the integrals in (\ref{4.3}) using
\begin{eqnarray}\label{4.4}
\int_{-\infty}^{\infty} dxdx'e^{-\alpha(x-x')^{2}}e^{-(x-A)^{2}}e^{-(x-B)^{2}}e^{-(x'-C)^{2}}e^{-(x'-D)^{2}}&=& \frac{\pi}{2\sqrt{\alpha+1}} e^{-\frac{1}{2}(A-B)^{2}}e^{-\frac{1}{2}(C-D)^{2}}
e^{-\frac{\alpha(A+B-C-D)^{2}}{4(\alpha+1)}},\nonumber\\
\end{eqnarray}
\noindent obtaining
\begin{eqnarray}\label{4.5}
_{k}\langle \ell_{i}|\hat \rho_{ss' k}(t)|\ell_{j}\rangle_{k}&=&\frac{1}{\frac{\lambda t}{\omega(k)}+1}e^{-\frac{\frac{\lambda t}{\omega(k)}}{2[1+\frac{\lambda t}{\omega(k)}]}dk| \tilde\chi_{i}(k)+\tilde\chi_{s}(k)-\tilde\chi_{j}(k)-\tilde\chi_{s'}(k)|^{2}}e^{-dk| \tilde\chi_{i}(k)-\tilde\chi_{s}(k)|^{2}}
e^{-dk| \tilde\chi_{j}(k)-\tilde\chi_{s'}(k)|^{2}},\nonumber\\
\langle \ell_{i}|\hat \rho_{ss'}(t)|\ell_{j}\rangle&=&Ke^{-\frac{\frac{\lambda t}{m}}{2[1+\frac{\lambda t}{m}]}\int_{0}^{\infty}dk| \tilde\chi_{i}(k)+\tilde\chi_{s}(k)-\tilde\chi_{j}(k)-\tilde\chi_{s'}(k)|^{2}}
e^{-\int_{0}^{\infty}dk| \tilde\chi_{i}(k)-\tilde\chi_{s}(k)|^{2}}
e^{-\int_{0}^{\infty}dk| \tilde\chi_{j}(k)-\tilde\chi_{s'}(k)|^{2}}\nonumber\\
&=&Ke^{-\frac{\frac{\lambda t}{m}}{4[1+\frac{\lambda t}{m}]}\int_{-\infty}^{\infty}dx [\chi_{i}(x)+\chi_{s}(x)-\chi_{j}(x)-\chi_{s'}(x)]^{2}}
e^{-\frac{1}{2}\int_{-\infty}^{\infty}dx[ \chi_{i}(x)-\chi_{s}(x)]^{2}}
e^{-\frac{1}{2}\int_{-\infty}^{\infty}dx[ \chi_{j}(x)-\chi_{s'}(x)]^{2}}\nonumber\\
\end{eqnarray}
\noindent where $K\equiv\prod_{k>0}\frac{1}{\frac{\lambda t}{\omega(k)}+1}$ (and, as mentioned earlier, replacing $\omega(k)$ by $m$ is a good approximation when multiplying $\tilde\chi(k)$). 
Using (\ref{3.8}), we rewrite $K$ as
\begin{equation}\label{4.6}
K=e^{-\sum_{k>0}\ln[\frac{\lambda t}{\omega(k)}+1]}=e^{-\frac{L}{2\pi}\int_{0}^{\infty}dk\ln[\frac{\lambda t}{\omega(k)}+1]}.             
\end{equation}
\noindent This will be discussed shortly.

From (\ref{4.5}), the elements of the density matrix are (using  $\int_{-\infty}^{\infty}dx\chi_{s}^{2}(x)=N$, $\int_{-\infty}^{\infty}dx\chi_{1}(x)\chi_{2}(x)\approx 0$ which follow from (\ref{2.3}), and setting $e^{-N}\approx 0$): 
\begin{eqnarray}\label{4.7} 
\langle \ell_{1}|\hat \rho(t)|\ell_{1}\rangle&=&\langle \ell_{2}|\hat \rho(t)|\ell_{2}\rangle=K\frac{1}{2}
\Big[1+e^{-2N}+2e^{-N\Big[1+\frac{\frac{\lambda t}{m}}{ 2[\frac{\lambda t}{m}     +1]}\Big]}\Big]
\approx K\frac{1}{2},\nonumber\\
\langle \ell_{1}|\hat \rho(t)|\ell_{2}\rangle&=&\langle \ell_{2}|\hat \rho(t)|\ell_{1}\rangle=K\frac{1}{2}
\Big[e^{-2N\frac{\frac{\lambda t}{m}}{ \frac{\lambda t}{m}     +1}}+e^{-2N}+2e^{-N\Big[1+\frac{\frac{\lambda t}{m}}{ 2[\frac{\lambda t}{m}     +1]}\Big]}\Big]
\approx K\frac{1}{2}e^{-2N\frac{\frac{\lambda t}{m}}{ \frac{\lambda t}{m}     +1}}.\nonumber\\
\end{eqnarray}

This is similar to non-relativistic  CSL collapse behavior, constant diagonal elements and decaying off-diagonal elements, here with exponent $\sim -\lambda Nt/m$ (although here the decay stops, 
but at negligible value $\sim e^{-2N}$).  

In addition however, there is the numerical factor $K = \prod_{k>0}\frac{1}{\frac{\lambda t}{\omega(k)}+1}=\langle \ell_{1}|\hat \rho(t)|\ell_{1}\rangle+\langle \ell_{2}|\hat \rho(t)|\ell_{2}\rangle$, the trace of $\hat\rho(t)$ with respect to the initial clump states. This is less than 1, and since  the trace of $\hat \rho(t)$ is 1, the trace of $\hat \rho(t)$ over all other states orthogonal to $|\ell_{1}\rangle,|\ell_{2}\rangle$ is $1-K$. These states 
are those for which the created particles are present, in addition to the initial clump states undergoing collapse behavior. As $t$ increases, these states of created particles come to dominate as, even for finite time, $K = 0$. This tells us that there is 0 probability of no particles created for $t>0$.

\section{Collapse Of A Superposition When $H\neq 0$}

The solution of Eq.(\ref{1.7}) for $\langle X_{k}|\langle p_{k}|\hat\rho_{ss' k} (t)|X'_{k}\rangle|p'_{k}\rangle$    is given in Appendix A, Eqs.(\ref{A12}),(\ref{A18}) with 
suitable identification of parameters, $X\rightarrow\sqrt{\omega dk}\tilde f_{R}(k), \; p\rightarrow\sqrt{\omega dk}\tilde f_{I}(k), \;\gamma_{1}\rightarrow\sqrt{2dk}\tilde\chi_{Rs}(k), \;
\gamma_{2}\rightarrow\sqrt{2dk}\tilde \chi_{Rs'}(k),\; \gamma'_{1}\rightarrow\sqrt{2dk}\tilde \chi_{Is}(k), \;
\gamma'_{2}\rightarrow\sqrt{2dk}\tilde\chi_{Is'}(k)$: 
\begin{eqnarray}\label{5.1}
 _{k}\langle f|\hat\rho_{ss' k} (t)|f'\rangle_{k}&=& \frac{2[1-S]}{\pi[1+S]}e^{-\frac{2Sdk\omega}{(1-S^{2})}|\tilde f(k)-\tilde f'(k)|^{2}} \nonumber\\
&&\cdot
e^{-dk\omega\frac{(1-S)}{(1+S)}\{|\tilde f(k) -\sqrt{\frac{2}{m}}\frac{\tilde \chi_{s}(k)-S\tilde \chi_{s'}(k)}{1-S}  |^{2}    +|\tilde f'(k) -\sqrt{\frac{2}{m}}\frac{\tilde \chi_{s'}(k)-S\tilde \chi_{s}(k)}{1-S}  |^{2} \}}
 e^{dk2\frac{S}{1-S}|\tilde \chi_{s}(k)-\tilde \chi_{s'}(k)|^{2}},\nonumber\\
\end{eqnarray}
\noindent with $S=\frac{\alpha}{1+\alpha}, \alpha = \frac{\lambda t}{2\omega(k)}$, so $S$ is a function of $k$.  

Eq.(\ref{5.1}) was derived as a good approximation for $t>>\frac{\hbar}{mc^{2}}$ which, for a neutron, is $\approx 10^{-23}$s, a negligible time on the scale of the collapse.  This removed some oscillating terms. 
The remainder of the oscillating terms were of the form $\tilde \chi_{s}(k)e^{\pm i\omega t}\approx \tilde \chi_{s}(k)e^{\pm im t}$ (since $\tilde \chi_{s}(k)$ is of negligible amplitude for 
relativistic $k$ values). Equation (\ref{5.1}) holds periodically, at the closely spaced times which are integer multiples of the period, so $e^{\pm im t}=1$  removes the remaining oscillating terms.  

Using (\ref{5.1}), we then can construct the complete density matrix in the $|f\rangle$ basis:
\begin{eqnarray}\label{5.2}
 \langle f|\hat\rho (t)|f'\rangle&=&\frac{1}{2}e^{-\int_{0}^{\infty}dk\frac{2S\omega}{(1-S^{2})}|\tilde f(k)-\tilde f'(k)|^{2}} \nonumber\\
&&\cdot
\sum_{ss'=1}^{2} e^{-\int_{0}^{\infty}dk\omega\frac{(1-S)}{(1+S)}\{|\tilde f(k) -\sqrt{\frac{2}{m}}\frac{\tilde \chi_{s}(k)-S\tilde \chi_{s'}(k)}{1-S}  |^{2}    +|\tilde f'(k) -\sqrt{\frac{2}{m}}\frac{\tilde \chi_{s'}(k)-S\tilde \chi_{s}(k)}{1-S}  |^{2} \}}
 e^{\int_{0}^{\infty}dk2\frac{S}{1-S}|\tilde \chi_{s}(k)-\tilde \chi_{s'}(k)|^{2}}\nonumber\\
 \end{eqnarray}
 \noindent absorbing the normalization factor $\frac{2[1-S]}{\pi[1+S]}$ in the functional integration element,  
$Df\equiv\prod_{k>0}  d\tilde f_{R}(k)d\tilde f_{I}(k)\frac{2[1-S]}{\pi[1+S]}\omega(k)dk$.

 This can be written in terms of $f(x)$ as follows. First, note that $S$ depends upon $\omega$, so write it explicitly $S(\omega)$ 
 and then make the approximation $S(\omega)\approx S(m)$, where $S$ multiplies  $\tilde \chi_{s}(k)$, since $\tilde \chi_{s}(k)$ has 
 non-relativistic momenta. This cannot be done where $S$ multiplies $\tilde f(k)$. Then, upon taking the Fourier transform, 
 note that $\omega$ becomes a differential operator, $\omega \rightarrow \hat{\omega} = [m^{2}-d^{2}/dx^{2}\big]^{1/4}$, and similarly $S(\omega)$ becomes a differential operator, $S(\omega)\rightarrow \hat{S} = S(\hat{\omega})$. The result is
 \begin{eqnarray}\label{5.3}
  \langle f|\hat\rho (t)|f'\rangle&\approx&\frac{1}{2}e^{-\int_{-\infty}^{\infty}dx\left\{\left[\frac{\hat{\omega}\hat{S}}{(1-\hat{S}^{2})}\right]^{1/2}\big[ f(x)- f'(x)\big]\right\}^2} \nonumber\\
&&\cdot
\sum_{ss'=1}^{2} e^{-\frac{1}{2}\int_{-\infty}^{\infty}dx \left\{\left[\hat{\omega}\frac{(1-\hat{S})}{(1+\hat{S})}\right]^{1/2}\big[ f(x) -\sqrt{\frac{2}{m}}\frac{\chi_{s}(x)-S(m)\chi_{s'}(x)}{1-S(m)} \big ]\right\}^{2}  } \nonumber \\
&&\qquad\cdot e^{-\frac{1}{2}\int_{-\infty}^{\infty}dx \left\{\left[\hat{\omega}\frac{(1-\hat{S})}{(1+\hat{S})}\right]^{1/2}
\big[ f'(x) -\sqrt{\frac{2}{m}}\frac{ \chi_{s'}(x)-S(m) \chi_{s}(x)}{1-S(m)}  \big] \right\}^{2} }\nonumber\\
&& \qquad\qquad\cdot e^{\int_{-\infty}^{\infty}dx\frac{S(m)}{1-S(m)}\big[ \chi_{s}(x)- \chi_{s'}(x)\big]^{2}},
\end{eqnarray}

Eqs.(\ref{5.1}),(\ref{5.2}),(\ref{5.3}) have the same general form as Eqs.(\ref{4.1}),(\ref{4.2}), except for the last exponential factor in (\ref{5.1}),(\ref{5.2}),(\ref{5.3}): this factor ensures that 
$Tr\hat\rho(t)=Tr\hat\rho(0)$ for the more complicated time dependence of these expressions.

The first (decaying) exponential factor in (\ref{5.3})  is large only if $ f(x)\approx  f'(x)$:  
this  characterizes the decay to eigenstates of $\hat\phi(x)$. For $s=s'$, the second two exponentials are large  if also $f(x)\approx f'(x)$, approaches the mass density distribution  
$\sqrt{\frac{2}{m}} \chi_{s}(x)$. Since these conditions can all be satisfied, their density matrix contribution can be large. 

 If $s\neq s'$, the density matrix must decay, since both of these exponentials cannot be large since $f(x),f'(x)$ have to approach different mass distributions when 
 $\chi_{s}(x)\neq\chi_{s'}(x)$. 
 
In fact, for short times,  using $1>>S\approx 0$, using Eq.(\ref{A13}),  expressions
(\ref{5.2}) and (\ref{5.3}) become \textit{identical} to (\ref{4.2}), the collapse expression when $H=0$.

For large times,  $1-S(\omega)\rightarrow \frac{2\omega}{\lambda t}$ and, using (\ref{A14}),  the limit of (\ref{5.2}), (\ref{5.3})  is  
\begin{eqnarray}\label{5.4}
\langle f|\hat\rho (t)|f'\rangle&=&\frac{1}{2}e^{-\frac{\lambda t}{2}\int_{0}^{\infty}dk|\tilde f(k)-\tilde f'(k)|^{2}}\sum_{ss'=1}^{2} 
e^{\int_{0}^{\infty}dk\sqrt{2m}\hbox{Re}[\tilde f(k)-\tilde f'(k)][\tilde \chi_{s}(k)-\tilde \chi_{s'}(k)]^{*}}
\nonumber\\
&&\cdot e^{-\frac{1}{\lambda t}\int_{0}^{\infty}dk\omega^{2} [ |\tilde f(k)|^{2}+|\tilde f'(k)|^{2}]}e^{-\int_{0}^{\infty}dk|\tilde \chi_{s}(k)-\tilde \chi_{s'}(k)|^{2}                 }
 \nonumber\\
&=&\frac{1}{2}e^{-\frac{\lambda t}{4}\int_{-\infty}^{\infty}dx\big[ f(x)- f'(x)\big]^{2}} 
\sum_{ss'=1}^{2} e^{\sqrt{\frac{m}{2}}\int_{-\infty}^{\infty}dx\big[ f(x)- f'(x)\big]\big[\chi_{s}(x)-\chi_{s'}(x)\big]} \nonumber\\
&&e^{-\frac{1}{2\lambda t}\int_{-\infty}^{\infty}dx\big[(\hat{\omega}f)^{2}(x)+(\hat{\omega}f')^{2}(x)\big]} 
e^{-\frac{1}{2}\int_{-\infty}^{\infty}dx\big[\chi_{s}(x)-\chi_{s'}(x)\big]^{2}}.
\end{eqnarray}
Note that the long time approximation is dependent on the mode. In order to apply it for all modes as we have done here, we are implicitly assuming that $f$ and $f'$ are chosen to be sufficiently smooth that $1-S(\omega)\sim \frac{2\omega}{\lambda t}$ is valid for all component modes.

For the diagonal terms, $s=s'$, clearly the first exponent dominates. There is no dependence upon the wave functions. The first exponential on the last 
line ensures the proper trace (were it not there, the trace would be infinite).
 
It is shown in Appendix A, Eq.(\ref{A15}), that this asymptotic behavior can be explained as each mode acting 
like a thermal density matrix $Ce^{-\frac{1}{k_{B}T}\hat H}$, where $\hat H=\omega [a^{\dagger}_{k}a_{k}+b^{\dagger}_{k}b_{k}]$, and with the 
temperature increasing with time, $\frac{\lambda t}{2\omega}=\frac{1}{e^{\frac{\omega}{k_{B}T}}-1}$. This is the eventual domination of the 
particle creation, caused by collapse to field eigenstates. 

For the off-diagonal terms, $s\neq s'$, really the same holds true. There is dependence on the wave function in the second exponential but, because the first exponential is only large if $f(x)\approx f'(x)$, this forces the second exponential towards the value 1,
negating the apparent dependence upon the wave function. (The wave function dependence in the last line has no time dependence, it is just the  
initial trace.)

\section{Concluding Remarks}
 In standard quantum theory, the state vector does not describe what 
actually happens, the occurrence of events.  If that is to be modified by CSL, so that description does occur, degrees of freedom have 
an increase of energy due to the CSL-induced narrowing of wave functions. 

In the classical physics lexicon, the world is made out of two kinds of things, particles and fields. In quantum theory, the two ideas 
merge, since each can be written in terms of the other. Particles have a finite number of degrees of freedom, while fields have an infinite number of degrees of freedom.  

Thus, 
the CSL-induced energy increase associated with particles is finite (and, in non-relativistic CSL small enough that it has not yet experimentally either been found or found not to exist), 
while the energy increase associated with fields is infinite, and therefore experimentally ruled out. If the CSL collapse mechanism is chosen by nature, 
it is a choice of particles over fields. The point of this paper has been to give the details of the road not taken.

\appendix

\section{Harmonic Oscillator and Collapse Generated by $\hat X$}

Here we give the solution of part of the Lindblad Eq.(\ref{1.6}), the harmonic oscillator when the collapse-generating operator is position:
\begin{eqnarray}\label{A1}
\frac{d}{dt}\hat\rho(t)
&=&-i\omega[\hat X^{2}+\frac{1}{4}\hat P^{2},\hat\rho(t)]-\frac{\lambda}{\omega}[\hat X, [\hat X, \hat\rho(t)]],            \nonumber\\                               
\end{eqnarray}
\noindent with the initial condition 
\begin{eqnarray}\label{A2}
\hat \rho(0)=e^{-\frac{1}{2}\gamma_{1}^{2}}e^{\gamma_{1}\hat a^{\dagger}}|0\rangle\langle 0|e^{\gamma_{2}\hat a}e^{-\frac{1}{2}\gamma_{2}^{2}},
\end{eqnarray}
\noindent  and  $\gamma_{1}, \gamma_{2}$ real. 

Note, $\hat \rho^{\dagger}(0)\neq\hat\rho(0)$, and  
$Tr\hat \rho(0)=e^{-\frac{1}{2}[\gamma_{1}-\gamma_{2}]^{2}}\neq 1$ (unless $\gamma_{1}=\gamma_{2}$): $\hat \rho(0)$ is \textit{part} of the initial 
density matrix utilized in section V, which of course is Hermitian and trace 1. Comparing the Hamiltonian to that of the usual harmonic oscillator, 
$\frac{1}{2}M\Omega^{2}\hat X^{2}+\frac{1}{2M}\hat P^{2}$, 
the relation is  $M=\frac{2}{\omega}, \Omega=\omega$, and so the annihilation operator is $\hat{a}=\sqrt{\frac{M\Omega}{2}} \hat X+i\sqrt{\frac{1}{2M\Omega}}\hat P=\hat X+i\frac{1}{2}\hat P$.

Eq.(\ref{A1}), expressed in terms of creation and annihilation operators,  is
\begin{eqnarray}\label{A3}
\frac{d}{dt}\hat \rho(t)&=&-i\omega [\hat a^{\dagger}\hat a,\hat \rho(t)]-\frac{\lambda}{4\omega}[\hat a^{\dagger}+\hat a,[\hat a^{\dagger}+\hat a,\hat\rho(t)].
\end{eqnarray}

We first will solve (\ref{A3}) for $\hat\rho(t)$ subject to the initial condititon (\ref{A2}), and then proceed to find $\langle X|\hat\rho(t)|X'\rangle$, which is utilized in Section V.

From (\ref{A3}), we find the equations for the expectation values of $\hat a,\hat a^{\dagger}, \hat a^{\dagger}\hat a, \hat a^{2}, \hat a^{\dagger 2}$, where e.g., 
$\overline{\hat a^{\dagger}\hat a(t)}\equiv Tr\hat a^{\dagger}\hat a\rho(t)$:
\begin{subequations}
\begin{eqnarray}\label{A4}
\frac{d}{dt}\overline{\hat a(t)}&=&-i\omega\hbox{Tr}\rho(t)[\hat a,\hat a^{\dagger}a]=-i\omega\overline{\hat a(t)}\hbox{ so }  \overline{\hat a(t)}=\gamma_{1}Tr\hat \rho(0) e^{-i\omega t},                             \label{A4a} \\
\overline{\hat a^{\dagger}(t)}&=&\gamma_{2}Tr\hat \rho(0)e^{i\omega t}                                  \label{A4b} \\ 
        \frac{d}{dt}\overline{\hat a^{\dagger}\hat a(t)}&=&-\frac{\lambda}{4\omega}\hbox{Tr}\hat\rho(t) [\hat a^{\dagger}+\hat a,[\hat a^{\dagger}+\hat a,\hat a^{\dagger}\hat a] =\frac{\lambda}{2\omega}Tr\hat \rho(0)    \hbox{ so }  \overline{\hat a^{\dagger}\hat a(t)}=[\frac{\lambda}{2\omega} t+\gamma_{1}\gamma_{2}]Tr\hat \rho(0)             \nonumber\\                         \label{A4c}            \\  
\frac{d}{dt}\overline{\hat a^{2}(t)}&=&- \frac{\lambda}{2\omega}Tr\hat \rho(0) -2i\omega\overline{\hat a^{2}(t)}   \hbox{ so }  \overline{\hat a^{2}(t)}= 
\Big[-\frac{\lambda}{2\omega}e^{-i\omega t} \frac{\sin\omega t}{\omega} +\gamma_{1}^{2}e^{-2i\omega t}\Big]Tr\hat \rho(0),                              \label{A4d} \\
   \overline{\hat a^{\dagger 2}(t)}&=& 
\Big[-\frac{\lambda}{2\omega}e^{i\omega t} \frac{\sin\omega t}{\omega} +\gamma_{2}^{2}e^{2i\omega t}\Big]Tr\hat \rho(0),                              \label{A4e} 
\end{eqnarray}
\end{subequations}

These expectation values detemine $\hat \rho(t)$, when we make the ansatz of the following quadratic form: 
\begin{eqnarray}\label{A5}
\hat \rho(t)&=&C(t)e^{R(t)\hat a^{\dagger 2}}e^{\beta_{1}(t)\hat a^{\dagger}}e^{S(t)\hat a_{L}^{\dagger}\hat a_{R}}|0\rangle\langle 0|e^{\beta_{2}^{*}(t)\hat a^{}}e^{R^{*}(t)\hat a^{2}}.
\end{eqnarray}

It follows from commuting the exponentials past $\hat a, \hat a^{\dagger}$ that 
\begin{eqnarray}\label{A6}
\hat a\hat \rho(t)&=&2R(t)\hat a^{\dagger}\hat \rho(t)+S(t)\hat \rho(t) \hat a+\beta_{1}(t)\hat \rho(t),\nonumber\\
\hat\rho \hat a^{\dagger}&=&2R^{*}(t)\hat \rho(t)\hat a+S(t)\hat a^{\dagger}\hat \rho(t) +\beta_{2}^{*}(t)\hat\rho(t)
\end{eqnarray}
\noindent (using $\hat a e^{S(t)\hat a_{L}^{\dagger}\hat a_{R}}|0\rangle\langle 0|=\sum_{n=0}^{\infty}\frac{S^{n}(t)}{n!}\hat a\hat a^{\dagger n}|0\rangle\langle 0|\hat a^{n}
=S(t) \sum_{n=1}^{\infty}\frac{S^{n-1}(t)}{(n-1)!}  \hat a^{\dagger (n-1)}|0\rangle\langle 0|\hat a^{n-1} \hat a=S(t)e^{S(t)\hat a_{L}^{\dagger}\hat a_{R}}|0\rangle\langle 0| \hat a$).
Then it follows from taking the trace of (\ref{A6}) and products of operators with (\ref{A6}) that
\begin{subequations}
\begin{eqnarray}
  \overline{\hat a(t)}&=&2R(t)\overline {\hat a^{\dagger}(t)}+S(t)\overline{\hat a(t)}+\beta_{1}(t)Tr\hat \rho(0),                           \label{A7a}                    \\
   \overline{\hat a^{\dagger}(t)}&=&2R^{*}(t)\overline {\hat a(t)}+S(t)\overline{\hat a^{\dagger}(t)}+\beta_{2}^{*}(t)Tr\hat \rho(0),          \label{A7b}                    \\
\overline{\hat a^{\dagger}\hat a(t)}&=&2R(t)\overline {\hat a^{\dagger 2}(t)}+S(t)[Tr\hat \rho(0)+\overline{\hat a^{\dagger}\hat a(t)}] +\beta_{1}(t)\overline{\hat a^{\dagger}(t)}                                 \nonumber                  \\ 
&=&2R^{*}(t)\overline {\hat a^{ 2}(t)}+S(t)[Tr\hat \rho(0)+\overline{\hat a^{\dagger}a(t)}] +\beta_{2}^{*}(t)\overline{\hat a(t)}\label{A7c}, \\
 \overline{\hat a^{2}(t)}&=&2R(t)[Tr\hat \rho(0)+\overline{\hat a^{\dagger}\hat a(t)}]   + S(t)\overline {\hat a^{2}(t)}+\beta_{1}(t)\overline{\hat a(t)} \label{A7d}, \\
\overline{\hat a^{\dagger 2}(t)}&=&2R^{*}(t)[Tr\hat \rho(0)+\overline{\hat a^{\dagger}\hat a(t)}]  + S(t)\overline {\hat a^{\dagger 2}(t)}+\beta_{2}^{*}(t)\overline{\hat a^{\dagger}(t)} \label{A7e}.
\end{eqnarray}
\end{subequations}

The solution of Eqs.(\ref{A7a}-\ref{A7e}), using (\ref{A4a}-\ref{A4e}), is
\begin{subequations}
\begin{eqnarray}
S(t)&=&1-\frac{\frac{\lambda}{2\omega}t+1}{[\frac{\lambda}{2\omega}t+1]^{2}-[\frac{\lambda}{2\omega^{2}}\sin\omega t]^{2}},      \label{A8a}                    \\
R(t)&=&\frac{-\frac{\lambda}{4\omega^{2}}e^{-i\omega t}\sin\omega t}{[\frac{\lambda}{2\omega}t+1]^{2}-[\frac{\lambda}{2\omega^{2}}\sin\omega t]^{2}} ,                             \label{A8b}                    \\ 
\beta_{1}(t)&=& \frac{[\frac{\lambda}{2\omega}t+1]\gamma_{1}e^{-i\omega t}+\lambda\gamma_{2}\frac{1}{2\omega^{2}}\sin\omega t}{[\frac{\lambda}{2\omega}t+1]^{2}-[\frac{\lambda}{2\omega^{2}}\sin\omega t]^{2}} ,                                                          \label{A8c}                    \\ 
\beta_{2}^{*}(t)&=& \frac{[\frac{\lambda}{2\omega}t+1]\gamma_{2}e^{i\omega t}+\lambda\gamma_{1}\frac{1}{2\omega^{2}}\sin\omega t}{[\frac{\lambda}{2\omega}t+1]^{2}-[\frac{\lambda}{2\omega^{2}}\sin\omega t]^{2}}.  \label{A8d}                                                                                                                                                                   
\end{eqnarray}
\end{subequations}

We now make some approximations. The characteristic time for the oscillations is $2\pi/\omega<2\pi/m\equiv\tau\approx 3\times 10^{-23}  $s. for a nucleon (a 
time scale very short compared to the collapse time).   
Therefore, after  say $t\approx 100\tau$, since $\frac{\lambda}{2\omega^{2}}\sin\omega t<\frac{ \lambda}{2\omega^{2}}<< \frac{\lambda}{2\omega}t $,   we may neglect the $\sin\omega t$ term, obtainiing 
\begin{subequations}
\begin{eqnarray}
S(t)&\approx&\frac{\frac{\lambda}{2\omega}t}{\frac{\lambda}{2\omega}t+1},      \label{A9a}                    \\
R(t)&\approx &0 ,                             \label{A9b}                    \\ 
\beta_{1}(t)&=&\frac{ \gamma_{1}e^{-i\omega t}}{\frac{\lambda}{2\omega}t+1} ,                                                          \label{A9c}                    \\ 
\beta_{2}^{*}(t)&=&\frac{ \gamma_{2}e^{i\omega t}}{\frac{\lambda}{2\omega}t+1}.   \label{A9d}                                                                                                                                                                   
\end{eqnarray}
\end{subequations}

Moreover, the exponentials $e^{\pm i\omega t}$ multiply $\gamma_{i}$. In our problem, $\gamma_{i} \sim \tilde \chi(k)$, which essentially vanishes unless $k\ll m$. 
So, we may set $e^{\pm i\omega t}\approx e^{\pm imt}$.  In that case, we may consider the solution only at times which are integer multiples of $\tau$,  since these 
are so closely spaced on the collapse dynamics time scale.  Therefore, so far we have  
\begin{equation}\label{A10}
   \hat \rho(t)\approx C(t)e^{\frac{ \gamma_{1}\hat a^{\dagger}}{\frac{\lambda t}{2\omega}t+1}}e^{\frac{\frac{\lambda t \hat a_{L}^{\dagger}\hat a_{R}}{2\omega}}{\frac{\lambda}{2\omega}t+1}}|0\rangle\langle 0|
  e^{\frac{ \gamma_{2}\hat a}{\frac{\lambda}{2\omega}t+1}}=C(t)e^{(1-S)\gamma_{1}\hat a^{\dagger}}e^{S\hat a_{L}^{\dagger}\hat a_{R}}|0\rangle\langle 0|
  e^{(1-S)\gamma_{2}\hat a}.
\end{equation}
It remains to find $C(t)$.  Since, according to (\ref{A3}), $\frac{d}{dt}Tr\hat\rho(t)=0$, it follows from (\ref{A10}) that
\begin{eqnarray}\label{A11}
0&=&\frac{1}{C(t)} \dot C(t) Tr\hat\rho(0)+\overline {\hat a^{\dagger}(t)}     \frac{d}{dt} \frac{ \gamma_{1}}{\frac{\lambda}{2\omega}t+1}   
+[Tr\hat\rho(0)+\overline {\hat a^{\dagger}a(t)} ]    \frac{d}{dt}\frac{\frac{\lambda t }{2\omega}}{\frac{\lambda}{2\omega}t+1}
+\overline {\hat a(t)}     \frac{d}{dt} \frac{ \gamma_{2}}{\frac{\lambda}{2\omega}t+1} \hbox{ or}\nonumber\\
0&=&\frac{1}{C(t)} \dot C(t)+\frac{\frac{\lambda  }{2\omega}}{\frac{\lambda}{2\omega}t+1}-\gamma_{1}\gamma_{2}\frac{\frac{\lambda  }{2\omega}}{[\frac{\lambda}{2\omega}t+1]^{2}}
\hbox{ with solution } \nonumber\\
C(t)&=&\frac{1}{\frac{\lambda}{2\omega}t+1}e^{\frac{\gamma_{1}\gamma_{2}\frac{\lambda }{2\omega}t}{\frac{\lambda}{2\omega}t+1}}
e^{-\frac{1}{2}[\gamma_{1}^{2}+\gamma_{2}^{2}]}=(1-S)e^{S\gamma_{1}\gamma_{2}}e^{-\frac{1}{2}[\gamma_{1}^{2}+\gamma_{2}^{2}]}       .            
\end{eqnarray}

\subsection{Density Matrix in the Position  Representation}

We now proceed to calculate the matrix element of (\ref{A10}), using $e^{S\hat a_{L}^{\dagger}\hat a_{R}}|0\rangle\langle 0|=\sum_{n=0}^{\infty}S^{n}|n\rangle\langle n|$, whose matrix 
elements in the position representation are given by a well-known identity 
involving Hermite polynomials \cite{Hermite}. We also  use the Campbell-Baker-Haussdorf theorem, obtaining:
\begin{eqnarray}\label{A12}
    \langle X|\hat\rho(t)|X'\rangle& =&C(t)e^{(1-S)\gamma_{1}[X-\frac{\partial}{2\partial X}]}e^{(1-S)\gamma_{2}[X'-\frac{\partial}{2\partial X'}]} |\sum_{n=0}^{\infty}S^{n}
  \langle X|n\rangle  \langle X'|n\rangle      \nonumber\\
  &=&C(t)e^{-\frac{1}{4}(1-S)^{2}\gamma_{1}^{2}}e^{-\frac{1}{4}(1-S)^{2}\gamma_{2}^{2}}e^{(1-S)\gamma_{1}X}e^{(1-S)\gamma_{2}X'}
  e^{-(1-S)\gamma_{1}\frac{\partial}{2\partial X}}  e^{-(1-S)\gamma_{2}\frac{\partial}{2\partial X'}}    \nonumber\\
  &&\cdot\frac{\sqrt{2}}{\sqrt{\pi[1-S^{2}]}}e^{-\frac{1-S}{2(1+S)}[X+X']^{2}}e^{-\frac{1+S}{2(1-S)}[X-X']^{2}}\nonumber\\
  &=&\sqrt{\frac{2[1-S]}{\pi[1+S]}}e^{S\gamma_{1}\gamma_{2}}e^{-\frac{1}{2}[\gamma_{1}^{2}+\gamma_{2}^{2}]} e^{-\frac{1}{4}(1-S)^{2}\gamma_{1}^{2}}e^{-\frac{1}{4}(1-S)^{2}\gamma_{2}^{2}}e^{(1-S)\gamma_{1}X}e^{(1-S)\gamma_{2}X'}\nonumber\\
 && \cdot e^{-\frac{1-S}{2(1+S)}[X+X'  -\frac{1-S}{2} (\gamma_{1}+\gamma_{2})   ]^{2}}e^{-\frac{1+S}{2(1-S)}[X-X' -\frac{1-S}{2} (\gamma_{1}-\gamma_{2}) ]^{2}}\nonumber\\
 &=&\sqrt{\frac{2[1-S]}{\pi[1+S]}}e^{-\frac{2S}{(1-S^{2})}[X-X']^{2}} e^{-\frac{1-S}{(1+S)}\{[X -\frac{\gamma_{1}-S\gamma_{2}}{1-S}   ]^{2}      +[X' -\frac{\gamma_{2}-S\gamma_{1}}{1-S}]^{2}\}}
 e^{\frac{S}{1-S}[\gamma_{1}-\gamma_{2}]^{2}}.
\end{eqnarray}

\subsection{Short Time and Long Time Limits}
For short times, $S\approx\frac{\lambda}{2\omega}t<<1$, (\ref{A12}) becomes
\begin{eqnarray}\label{A13}
    \langle X|\hat\rho(t)|X'\rangle& \approx&
\sqrt{\frac{2}{\pi}}e^{-\frac{\lambda t}{\omega}[X-X']^{2}} e^{-[X -\gamma_{1}]^{2}}  e^{-[X' -\gamma_{2}]^{2}} .
\end{eqnarray}
\noindent It is consistent to neglect the exponent in the last factor in (\ref{A12}), $e^{\frac{S}{1-S}[\gamma_{1}-\gamma_{2}]^{2}}\approx e^{-\frac{\lambda t}{2\omega}[\gamma_{1}-\gamma_{2}]^{2}}\approx 1$ 
to accompany the approximation $S\gamma_{i}<<\gamma_{i}$, so that, even in this 
approximation, the property that 
the trace is unchanged is preserved, $Tr\hat\rho(t)=Tr\hat\rho(0)=e^{-\frac{1}{2}[\gamma_{1}-\gamma_{2}]^{2}}.$ 

For long times, using $1-S\approx \frac{2\omega}{\lambda t}$, (\ref{A12}) becomes
\begin{eqnarray}\label{A14}
    \langle X|\hat\rho(t)|X'\rangle\approx\sqrt{\frac{2\omega}{\pi\lambda t}} e^{-\frac{\lambda t}{2\omega}[X-X']^{2}} e^{[X-X'][\gamma_{1}-\gamma_{2}]} e^{-\frac{\omega}{\lambda t}[X^{2}+X'^{2}]}e^{-\frac{1}{2}[\gamma_{1}-\gamma_{2}]^{2}}. 
\end{eqnarray}
The first exponential dominates: it is large only if $|X-X'|^{2}\sim \frac{2\omega}{\lambda t}$, in which case the second and third exponentials approach 1 as $t\rightarrow\infty$. The third and fourth exponentials are there to preserve the property that the trace is unchanged. 

To understand this behavior better, return to Eq.(\ref{A10}), where the $(1-S)\gamma_{i}\rightarrow \gamma_{i}\frac{2\omega}{\lambda t}$ dependence makes it clear that 
the $\gamma_{i}$ dependence is asymptotically negligible. Then, concentrate on the Trace 1 remainder: 
\begin{eqnarray}\label{A15}
\hat\rho(t)&=& (1-S)e^{S\hat a_{L}^{\dagger}\hat a_{R}}|0\rangle\langle 0|   =(1-S)\sum_{n=1}^{\infty}S^{n}|n\rangle\langle n|\nonumber\\
&=& (1-S)\sum_{n=1}^{\infty}e^{n\ln S} |n\rangle\langle n|= (1-S)e^{\hat a^{\dagger}\hat a\ln S}\sum_{n=1}^{\infty} |n\rangle\langle n|  \nonumber\\
&=&(1-S)e^{\hat a^{\dagger}\hat a\ln S}.                  
\end{eqnarray}
\noindent This is a thermal density matrix, with identification of $S=e^{-\frac{\omega}{k_{B}T}}$, with temperature $T$ (and $k_{B}$ the Boltzmann constant) and Hamiltonian 
 $=\omega\hat a^{\dagger}\hat a$. The mean particle number is $Tr \hat a^{\dagger}\hat a\rho(t)=\frac{S}{1-S}= \frac{\lambda t}{2\omega}=\frac{1}{e^{\frac{\omega}{k_{B}T}}-1}$.       
 So, the particle number increase with increasing time is the same as the occupation number increase with increasing temperature, of a harmonic oscillator in a thermal bath, always in thermal equilibrium.

\subsection{Harmonic Oscillator and Collapse Generated by $\hat p$ }

The other harmonic oscillator problem in (\ref{1.6}),
 \begin{eqnarray}\label{A16}
\frac{d}{dt}\hat\rho(t)
&=& -i\omega[ \frac{1}{4}\hat x^{2}+\hat p^{2},\hat\rho(t)] -\frac{\lambda}{\omega} [\hat p, [\hat p, \hat\rho(t)]]\Big\}\nonumber\\
&=&  -i\omega [\hat a^{\dagger}\hat a, \rho(t)]-\frac{\lambda}{4\omega}[\hat a^{\dagger}+\hat a,[\hat a^{\dagger}+\hat a,\hat\rho(t)],                             
\end{eqnarray}
 subject to the initial condition
\begin{eqnarray}\label{A17}
\hat \rho(0)=e^{-\frac{1}{2}\gamma_{1}^{'2}}e^{\gamma'_{1}\hat a^{\dagger}}|0\rangle\langle 0|e^{\gamma'_{2}\hat a}e^{-\frac{1}{2}\gamma_{2}^{'2}},
\end{eqnarray}
\noindent has precisely the same solution:
\begin{eqnarray}\label{A18}
    \langle p|\rho(t)|p'\rangle
 &=&\sqrt{\frac{2[1-S]}{\pi[1+S]}}e^{-\frac{2S}{(1-S^{2})}[p -p']^{2}} e^{-\frac{1-S}{(1+S)}\{[p -\frac{\gamma'_{1}-S\gamma'_{2}}{1-S}   ]^{2}      +[p' -\frac{\gamma'_{2}-S\gamma'_{1}}{1-S}]^{2\}}}
 e^{\frac{S}{1-S}[\gamma'_{1}-\gamma'_{2}]^{2}}.
\end{eqnarray}



\begin{thebibliography}{99}
\bibitem{Hermite} The identity (e.g., in the last section of Wikipedia's article on Hermite polynomials) is
\begin{equation}
                                \sum_{n=0}^{\infty}S^{n}\psi_{n}(x)\psi_{n}(y)=\frac{1}{\sqrt{\pi[1-S^{2}]}}\Big[e^{-\frac{1-S}{1+S}\frac{(x+y)^{2}}{4}}
                                +e^{-\frac{1+S}{1-S}\frac{(x-y)^{2}}{4}}\Big]
\end{equation}
\noindent where $\psi_{n}(x)$ is the usual harmonic oscillator wave function.  However, the usual annihilation operator is 
$a=\frac{1}{\sqrt{2}}[\hat X+i\hat P]$ whereas the one used here is $a=[\hat X+i\frac{1}{2}\hat P]$. This means that the 
wave functions $\langle X|n\rangle=2^{1/4}\psi_{n}(\sqrt{2}X)$.
\end{thebibliography}
\end{document}